\documentclass[pdflatex,sn-nature, iicol]{sn-jnl}

\usepackage{graphicx}%
\usepackage{multirow}%
\usepackage{amsmath,amssymb,amsfonts}%
\usepackage{amsthm}%
\usepackage{mathrsfs}%
\usepackage[title]{appendix}%
\usepackage{xcolor}%
\usepackage{textcomp}%
\usepackage{manyfoot}%
\usepackage{booktabs}%
\usepackage{algorithm}%
\usepackage{algorithmicx}%
\usepackage{algpseudocode}%
\usepackage{listings}%

\usepackage[capitalise]{cleveref}

\renewcommand{\Cref}{\cref}
\newcommand{\hide}[1]{}

\usepackage{subcaption}
\usepackage{siunitx}

\usepackage[version=4]{mhchem}

\theoremstyle{thmstyleone}%
\theoremstyle{thmstyletwo}%

\theoremstyle{thmstylethree}%

\begin{document}


\title[Article Title]{Fully analogue in-memory neural computing via quantum tunneling effect}

\author[1]{\fnm{Songyuan} \sur{Li}}
\equalcont{These authors contributed equally to this work.}

\author[1]{\fnm{Teng} \sur{Wang}}
\equalcont{These authors contributed equally to this work.}

\author[1]{\fnm{Jinrong} \sur{Tang}}
\author[1]{\fnm{Ruiqi} \sur{Liu}}
\author[1]{\fnm{Haoyu} \sur{Li}}

\author[2]{\fnm{Yuyao} \sur{Lu}}
\author[2]{\fnm{Feng} \sur{Xu}}
\author[2]{\fnm{Bin} \sur{Gao}}
\author[3,4]{\fnm{Can} \sur{Xie}}

\author*[1]{\fnm{Xiangwei} \sur{Zhu}}\email{zhuxw666@mail.sysu.edu.cn}

\affil*[1]{\orgdiv{School of Electronics and Communication Engineering}, \orgname{Sun Yat-sen University}, \orgaddress{ \city{Shenzhen}, \postcode{518107},  \country{China}}}

\affil[2]{\orgdiv{School of Integrated Circuits, Innovation Center for Future Chips}, \orgname{Tsinghua University}, \orgaddress{ \city{Beijing}, \postcode{100084}, \country{China}}}
 
\affil[3]{\orgdiv{Institute of Quantum Sensing}, \orgname{Zhejiang University}, \orgaddress{\city{Hangzhou}, \postcode{310027},  \country{China}}}

\affil[4]{\orgdiv{College of Life and Environmental Sciences}, \orgname{Hangzhou Normal University}, \orgaddress{\city{Hangzhou}, \postcode{311121},  \country{China}}}


\abstract{

Fully analogue neural computation requires hardware that can implement both linear and nonlinear transformations without digital assistance. While analogue in-memory computing efficiently realizes matrix-vector multiplication, the absence of learnable analogue nonlinearities remains a central bottleneck. Here we introduce KANalogue, a fully analogue realization of Kolmogorov-Arnold Networks (KANs) that instantiates univariate basis functions directly using negative-differential-resistance (NDR) devices. By mapping the intrinsic current-voltage characteristics of NDR devices to learnable coordinate-wise nonlinear functions, KANalogue embeds function approximation into device physics while preserving a fully analogue signal path. Using cold-metal tunnel diodes as a representative platform, we construct diverse nonlinear bases and combine them through crossbar-based analogue summation. Experiments on MNIST, FashionMNIST, and CIFAR-10 demonstrate that KANalogue achieves competitive accuracy with substantially fewer parameters and higher crossbar node efficiency than analogue MLPs, while approaching the performance of digital KANs under strict hardware constraints. The framework is not limited to a specific device technology and naturally generalizes to a broad class of NDR devices. These results establish a device-grounded route toward scalable, energy-efficient, fully analogue neural networks.

%
}




\maketitle


\begin{figure*}[tb]
	\centering
	\includegraphics[width=\textwidth]{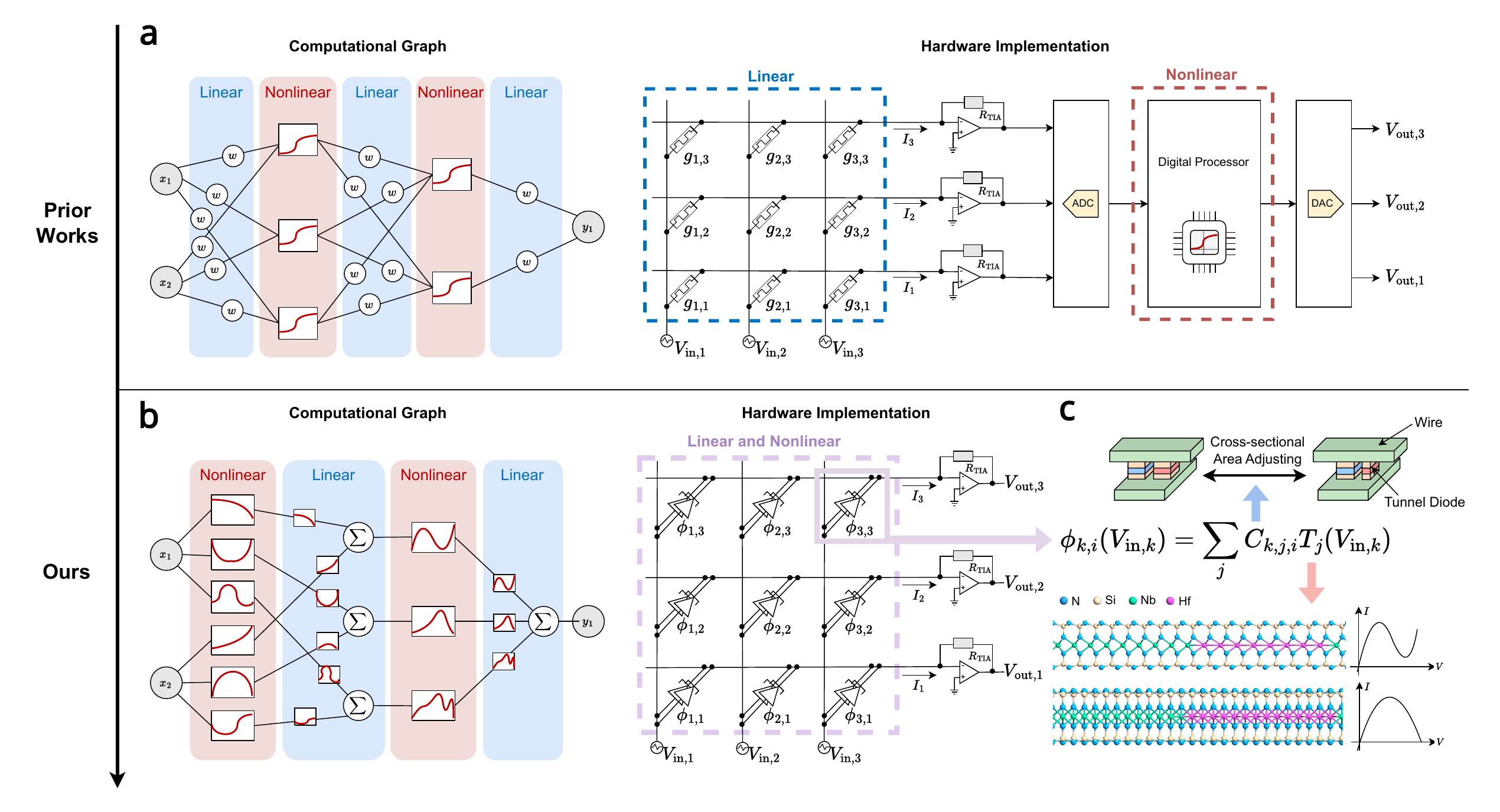}
	\captionsetup{justification=raggedright, singlelinecheck=false}
	\caption{
\textbf{Comparison between prior analogue neural architectures and the proposed KANalogue framework.}
\textbf{a, Prior works:} Conventional analogue accelerators implement only the linear operations of neural networks in crossbar arrays, while nonlinear activations remain fixed and must be computed digitally after analogue-to-digital conversion. This hybrid pipeline limits efficiency and prevents fully analogue neural computation.  
\textbf{b, This work:} KANalogue adopts the KAN architecture, in which learnable univariate functions reside on edges and additive mixing occurs at nodes. This structure naturally maps to hardware: both linear summation and nonlinear basis evaluations are implemented directly in the crossbar array, eliminating the digital nonlinear-processing stage.  
\textbf{c, Device-level realization:} Learnable activation functions are instantiated using cold-metal tunnel diodes (CMTDs). Distinct device geometries yield diverse nonlinear I--V characteristics, which serve as tunable basis functions \(T_{k,j,i}(x)\). Weighted combinations by cross-sectional area adjusting of these device-derived basis functions form the analogue edge mappings \(\phi_{k,i}\). This device-to-function correspondence enables a fully analogue computational path grounded in physical nonlinearities.
} \label{fig:intro}
\end{figure*}

The growing demand for efficient machine learning on edge devices has intensified the search for alternatives to conventional digital architectures, which often struggle with power and latency constraints~\cite{ambrogio2023analog, wen2024fusion}. Neuromorphic computing offer a compelling pathway, exploiting device physics to achieve orders-of-magnitude gains in energy efficiency and computational density~\cite{mead1990neuromorphic,kudithipudi2025neuromorphic}. Recent advances in analogue in-memory computing (AIMC) based on resistive RAM~(RRAM) and phase-change memory~(PCM) have demonstrated near-software-level accuracy for large-scale matrix-vector multiplication~\cite{buchel2024kernel, huang2024memristor, zhang2021computing,LeGallo2023,Rasch2024,aguirre2024hardware,Wu2024,lanza2025growing,wang2023echo}. By leveraging Ohm's law and Kirchhoff's current law, AIMC arrays physically co-locate memory and computation to perform these linear operations \emph{in situ}, directly alleviating the von Neumann bottleneck that arises from shuttling data between separate memory and processing units~\cite{sebastian2020memory,xi2020memory,seok2024beyond}. Yet neural networks are not purely linear systems: they rely on alternating linear and nonlinear transformations~(\cref{fig:intro}a). A critical obstacle remains because most analogue accelerators still delegate nonlinear activation functions to digital circuits, breaking the vision of fully analogue neural pipelines and limiting overall efficiency~\cite{buchel2024kernel, Yang2025}.

Kolmogorov-Arnold Networks (KANs) provide an opportunity to address this gap. Inspired by the Kolmogorov-Arnold representation theorem~\cite{kolmogorov1957representations}, KANs replace edge weights with learnable coordinate-wise basis functions. This structural distinction shifts expressive capacity from dense matrix multiplication to univariate nonlinear transformations, enabling compact models that differ fundamentally from traditional multilayer perceptrons~\cite{liu2025kan}. Crucially, it also makes KANs particularly suitable for analogue realization: each basis function can, in principle, be instantiated directly by a physical nonlinear device. Combined with their reduced parameter counts compared to conventional networks, KANs have the potential to offer a smaller hardware footprint and lower energy demands. Recent variants have been applied to symbolic regression~\cite{sidharth2024}, modeling of physical systems~\cite{aghaei2025fkan,bozorgzadeh2024wavelet,toscano2024kkan}, remote-sensing image classification~\cite{cheon2024remote}, and molecular property prediction~\cite{li2025kolmogorov}. Despite these advantages, KANs have so far been realized only in digital or hybrid platforms~\cite{huang2025hardware,peng2024photonic}, leaving their analogue potential largely unexplored.

Nonlinear devices with negative differential resistance (NDR) offer a promising route to close this gap. Esaki diodes~\cite{esaki1958new}, resonant tunneling diodes~\cite{tsu1973tunneling,chang1974resonant}, and recently designed cold-metal devices~\cite{theoretical,Bodewei2024} naturally exhibit nonlinear current-voltage characteristics that resemble activation functions. Prior work has demonstrated the use of NDR devices to implement spiking neurons~\cite{Donati2024}, bistable logic gates~\cite{McNaughton2025}, or to emulate software-defined nonlinearities in memristor crossbars~\cite{Yang2025}. However, integrating device-level nonlinearities directly into function-based architectures has yet to be realized.

Here we introduce KANalogue, a framework that implements Kolmogorov-Arnold Networks using NDR devices as physical instantiations of univariate basis functions~(\cref{fig:intro}b).  NDR devices exhibit inherently nonlinear, non-monotonic current-voltage characteristics that provide rich expressive capability. Moreover, multiple NDR elements can be composed to realize more complex nonlinear functions, further extending the expressive capacity of analogue nodes~(\cref{fig:intro}c). This property allows KAN nodes to approximate complex mappings with fewer units, improving node efficiency and enabling more compact network realizations. By mapping these device characteristics to the functional components of KANs, KANalogue establishes a direct bridge between quantum electronic behavior and neural computation. Through simulation and surrogate modeling based on cold-metal tunnel diodes, we demonstrate that NDR-enabled KANs achieve competitive accuracy on standard benchmarks while reducing parameter counts and eliminating digital nonlinearity computation. When scaled to larger systems, such analogue implementations naturally exhibit statistical robustness: variations and noise across individual devices are averaged out through distributed computation, imparting a degree of error tolerance to the overall network. This work integrates theoretical function approximation with device-level physics, pointing toward scalable, fully analogue neural networks grounded in emerging, manufacturable device technologies.

\section*{Results}
\subsection*{KANalogue: device-grounded KANs}
\label{sec:KANframework}
\begin{figure*}[htbp]
	\centering
	\includegraphics[width=1\textwidth]{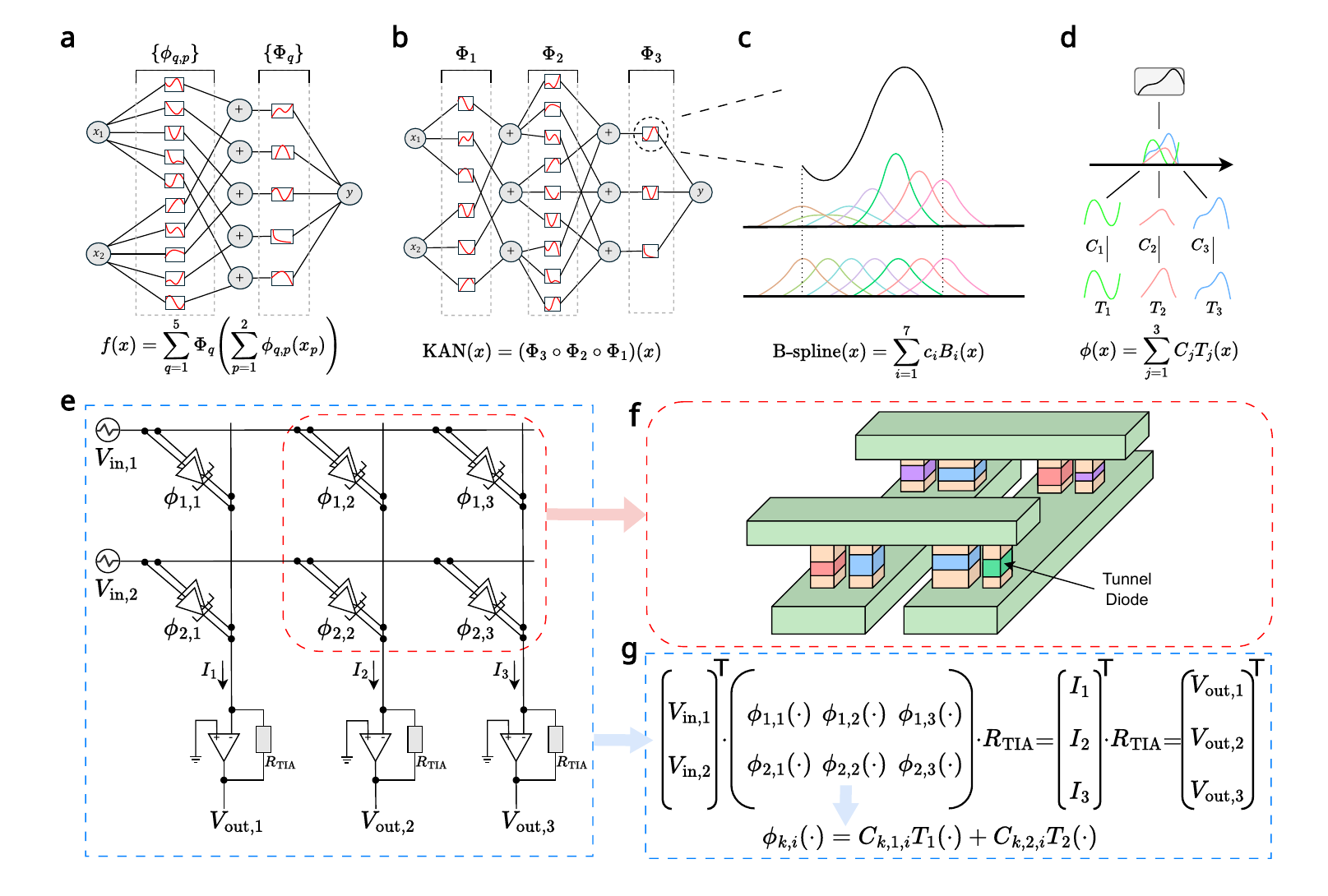}
	\captionsetup{justification=raggedright, singlelinecheck=false}
	\vspace{-2em}
	\caption{\small
	\textbf{KANalogue framework.}
	\textbf{a,} KART expresses any continuous multivariate function as a finite superposition of univariate functions. This structure can be viewed as a two-layer network in which univariate mappings reside on the edges and additive mixing occurs at the nodes.
	\textbf{b,} KANs generalize this construction into a deep architecture by stacking multiple such layers, with each edge carrying a learnable univariate function.
	\textbf{c,} In original KANs, each univariate function is parameterized by a B-spline expansion with trainable coefficients.
	\textbf{d,} KANalogue transfers these basis functions into hardware by replacing digital B-splines with nonlinear I--V characteristics of NDR devices. 
	\textbf{e,}  Analogue crossbar realization of a KANalogue layer. Each edge combines multiple NDR devices and each node perform current summation governed by Kirchhoff's current law.
	\textbf{f,} Schematic illustration of a $2\times 2$ physical crossbar, showing how device geometry encodes effective conductance.
	\textbf{g,} Matrix-form representation of a KANalogue layer. Nonlinear device responses form learnable basis functions on each edge, while transimpedance amplification~(TIAs) converts summed currents into output voltages.
	} \label{fig:framework}
\end{figure*}

KANalogue realizes KANs directly in hardware by mapping their univariate basis functions to the nonlinear I--V responses of NDR devices. 
The construction is grounded in the Kolmogorov-Arnold representation theorem (KART)~\cite{kolmogorov1957representations}, which states that any continuous function $f:\mathbb{R}^n \to \mathbb{R}$ can be written as a finite superposition of univariate functions:
\begin{equation}
	f(x_1,\ldots,x_n)
	= \sum_{q=1}^{2n+1} \Phi_q\!\left( \sum_{p=1}^{n} \phi_{q,p}(x_p) \right),
\end{equation}
where $\Phi_q(\cdot)$ and $\phi_{q,p}(\cdot)$ are univariate mappings.
This decomposition can be viewed as a two-layer network in which univariate functions live on the \emph{edges} and summations at the \emph{nodes} implement additive mixing (Fig.~\ref{fig:framework}a).

KANs generalize this idea by discarding the specific bound $2n{+}1$ and stacking multiple such layers into a deep architecture (Fig.~\ref{fig:framework}b). 
Each edge is associated with a learnable univariate function, which in the original formulation is parameterized by B-spline expansions
\begin{equation}\label{eq:b-spline}
	\mathrm{B\text{-}spline}(x) = \sum_i c_i B_i(x),
\end{equation}
where $B_i(x)$ are spline basis functions and $c_i$ are trainable coefficients (Fig.~\ref{fig:framework}c). 
Learning these coefficients endows KANs with strong approximation power while retaining a structured representation rooted in KART.

KANalogue transfers this functional structure into the analogue domain. 
Instead of implementing $B_i(x)$ digitally, each basis element is instantiated by the measured or simulated I--V characteristic of an NDR device (Fig.~\ref{fig:framework}d). 
A KAN edge is thus realized as a small analogue circuit that combines several device-derived basis functions, while a KAN node corresponds to current summation governed by Kirchhoff’s current law (Fig.~\ref{fig:framework}e-g). 
By embedding the basis functions directly in device physics, KANalogue preserves the expressivity of KANs while enabling an all-analogue signal path from input to output.

%

\subsection*{Fully analogue KAN architecture}
\label{sec:implement}


Realizing KANs entirely in hardware requires implementing every component of a KAN layer—including its nonlinear basis functions, linear mixing operations, and signal-range regulation—in the analogue domain. In KANalogue, the nonlinear univariate transformations are provided by NDR devices whose I--V characteristics act as physically realizable basis functions. These device-derived nonlinearities are then linearly combined through analogue crossbar arrays that naturally perform vector-matrix multiplication via conductance-weighted summation. To ensure that all signals remain within the physical operating range of the hardware, the analogue pipeline further incorporates a hard-clip mechanism and a normalization strategy whose parameters are fused into the layer weights after training. 

\begin{figure*}[htbp]
	\centering
	\includegraphics[width=1\textwidth]{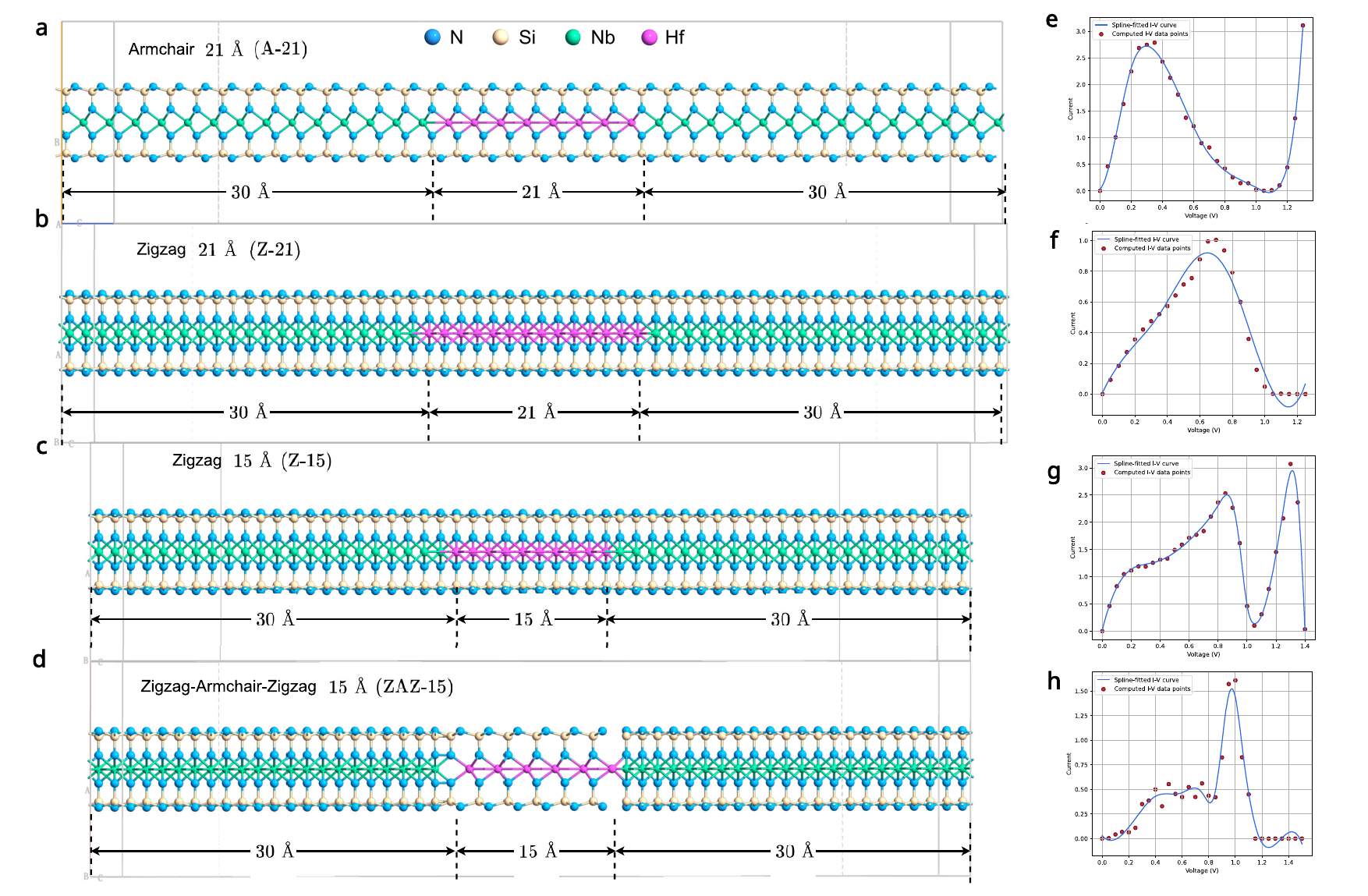}
	\captionsetup{justification=raggedright, singlelinecheck=false}
	\caption{ \small
\textbf{CMTD structures and their fitted nonlinear I--V characteristics.}
\textbf{a,} Armchair-oriented \ce{NbSi2N4}/\ce{HfSi2N4} heterostructure with a \SI{21}{\angstrom} tunnel barrier.  
\textbf{b,} Zigzag-oriented heterostructure with the same \SI{21}{\angstrom} barrier length.  
\textbf{c,} Zigzag-oriented device with a shorter \SI{15}{\angstrom} barrier, highlighting the effect of barrier thickness on tunneling behaviour.  
\textbf{d,} Hybrid zigzag–armchair–zigzag configuration with a \SI{15}{\angstrom} barrier region.  
\textbf{e-h,} Corresponding I--V characteristics obtained from DFT--NEGF simulations (dots) and their smoothing-spline fits (solid curves), which serve as analogue nonlinear basis functions in KANalogue. The variations across geometry and orientation illustrate the device-level tunability of the nonlinear transformations.}

	\label{fig:cmtd}
\end{figure*}

\subsubsection*{Device-Derived Nonlinear Basis Functions}
A central requirement for realizing KANs in the analogue domain is the availability of configurable univariate nonlinear functions that can serve as basis transformations. NDR devices naturally satisfy this requirement: their nonmonotonic current-voltage (I--V) characteristics provide smooth and tunable nonlinear responses that can be shaped through material choice and device geometry. While several device families exhibit NDR behaviour—including Esaki diodes, resonant-tunneling structures, and interband tunneling devices—we focus on cold-metal tunnel diodes (CMTDs), which offer a broader design space and greater material-level programmability~\cite{sasioglu2022proposal,zhang2024high}.

CMTDs exhibit strong geometry-dependent nonlinear behaviour: variations in atomic arrangement, barrier thickness, and electrode configuration directly modify the peak and valley currents, the onset of the NDR window, and the cold-tunneling slope. Using first-principles density-functional-theory-based nonequilibrium Green's function (DFT-NEGF) simulations (Methods), we generate I–V characteristics for \ce{NbSi2N4}/\ce{HfSi2N4} lateral heterostructures across a range of barrier lengths and crystallographic orientations (Fig.~\ref{fig:cmtd}). This process yields a physically tunable, device-level family of nonlinear transformations~\cite{Bodewei2024}.

To incorporate these device responses into KANalogue, each simulated I--V characteristic is fitted using a smoothing spline, producing a continuous nonlinear mapping that is robust to numerical noise while preserving the intrinsic device behaviour. We denote these fitted nonlinear functions by $T_j(\cdot)$ and construct a basis set $\{T_j(\cdot)\}_{j=1}^{d_{\text{basis}}}$ by selecting $d_{\text{basis}}$ CMTDs whose responses are non-redundant over the operating voltage range. This device-derived basis replaces the conventional B-spline basis used in digital KAN implementations (cf.~\cref{eq:b-spline}).


 \subsubsection*{Linear mixing via analogue crossbars}

Once the device-derived nonlinear basis functions are defined, the remaining operation of a KAN layer is linear: the nonlinear responses generated across input dimensions and basis indices must be aggregated to form each output channel. In KANalogue, this aggregation is performed entirely in the analogue domain and constitutes the second core component of the architecture.

Given an input vector $\mathbf{x} \in \mathbb{R}^{d_{\text{in}}}$, each basis function $T_j(\cdot)$ is evaluated elementwise on the input components $x_k$, producing a set of nonlinear features $\{T_j(x_k)\}$. For a given output channel $i$, the contribution from input dimension $k$ is expressed as a linear combination of these basis responses,
\begin{equation}\label{eq:phi}
\phi_{k,i}(x_k) = \sum_{j=1}^{d_{\text{basis}}} C_{k,j,i}\, T_j(x_k),
\end{equation}
where $C_{k,j,i}$ are trainable coefficients. Summing these contributions across all input dimensions yields the output of the analogue KAN layer,
\begin{equation}
y_i = \sum_{k=1}^{d_\text{in}}\phi_{k,i}(x_k),
\label{eq:kan_layer}
\end{equation}
which corresponds to a linear transformation applied to nonlinear, device-generated features.

From a hardware perspective, this operation is naturally realized by an analogue crossbar array~(\cref{fig:framework}e). Input voltages
$\mathbf{V}_{\text{in}} = [V_{\text{in},1}, \ldots, V_{\text{in},d_{\text{in}}}]^{\mathsf T}$
encode the input components, while device-level nonlinearities generate current responses proportional to $T_j(V_{\text{in},k})$. The coefficients $C_{k,j,i}$ are implemented through conductance scaling, for example via device geometry or effective cross-sectional area~(\cref{fig:framework}f). Along each column of the crossbar, Kirchhoff's current law performs the summation of these contributions, producing output currents that are subsequently converted to voltages by transimpedance amplifiers.

In matrix form, the analogue linear stage can be written as
\begin{equation}
\mathbf{V}_{\text{out}}
= R_{\text{TIA}}\,
\mathbf{V}_{\text{in}}^{\mathsf T}\,
\underbrace{
\begin{bmatrix}
\phi_{1,1}(\cdot) & \phi_{1,2}(\cdot) & \cdots & \phi_{1,d_{\text{out}}}(\cdot) \\
\phi_{2,1}(\cdot) & \phi_{2,2}(\cdot) & \cdots & \phi_{2,d_{\text{out}}}(\cdot) \\
\vdots & \vdots & \ddots & \vdots \\
\phi_{d_{\text{in}},1}(\cdot) &
\phi_{d_{\text{in}},2}(\cdot) & \cdots &
\phi_{d_{\text{in}},d_{\text{out}}}(\cdot)
\end{bmatrix}
}_{\boldsymbol{\Phi}},
\label{eq:vin_vout_phi_dot}
\end{equation}
where $\boldsymbol{\Phi}$ denotes an input-dependent effective conductance matrix, whose entries encode the aggregated nonlinear response from each input line to each output column as in \cref{eq:phi}, and $R_{\text{TIA}}$ is the transimpedance gain converting current to voltage (Fig.~\ref{fig:framework}g). This formulation highlights how nonlinear transformation and linear aggregation are jointly realized through analogue device physics and circuit-level summation.


%
%
%

\subsubsection*{Signal conditioning and range control}

We apply KANalogue to image classification as a concrete testbed. 
Figure~\ref{fig:pipeline_result}a summarizes the overall pipeline: high-dimensional inputs are first mapped into the operating range of the devices, then transformed by stacks of KANalogue layers, and finally read out by a linear classifier.  
A complete analogue implementation requires that all intermediate signals remain within the safe operating range of the hardware. Because analogue voltages and currents cannot grow unbounded, KANalogue incorporates a lightweight signal-conditioning stage that stabilizes the analogue forward path without introducing any digital computation. This stage combines an analogue hard-clip that limits signal amplitude with a normalization strategy used only during training and later fused into the network weights.

The outputs of each KAN layer are continuous analogue signals produced by the superposition of device-based nonlinearities. To prevent these signals from exceeding the dynamic range of subsequent layers or saturating the crossbar, we apply a hard-clip that bounds the voltage to a predefined interval. In hardware, this behaviour can be implemented using a limiting amplifier or rail-to-rail saturation stage, which restricts the output automatically as the signal approaches the supply rails. The clipper imposes no quantization and therefore preserves the continuous analogue nature of the computation.

During training, Batch Normalization (BatchNorm) improves optimization stability by regulating signal variance. After training, however, the affine parameters of BatchNorm are folded into the coefficient tensor \(C\), removing the need for explicit normalization circuitry at inference. This fusion ensures that the deployed network operates entirely in the analogue domain, with no per-sample normalization.

Together, the clipper and fused normalization produce a well-conditioned signal that remains within device-safe bounds and is compatible with the analogue nonlinearities and crossbar mixing. As a result, each KAN layer forms a fully analogue pipeline—device-derived nonlinear basis functions, crossbar-based linear combination, amplitude regulation—without any digital post-processing.

\subsection*{Expressive role of basis-function diversity}
\label{sec:ablation}

KANalogue derives its nonlinear expressivity from a small set of device-defined basis functions, each corresponding to a distinct CMTD I--V characteristic. Because different CMTD geometries give rise to markedly different nonlinear profiles, the choice and combination of basis functions plays a central role in determining the expressive capacity of the model.

To examine how basis diversity influences performance, we systematically evaluate KANalogue models constructed from different combinations of CMTD-based basis functions.  We consider four distinct spline-fitted CMTD responses~(\cref{fig:cmtd}), which give rise to multiple combinations of basis functions. We refer to the number of basis functions used in a given configuration as the \emph{basis dimension}. Experiments are conducted on MNIST, FashionMNIST, and CIFAR-10 using dataset-appropriate network architectures. Because the number of trainable parameters scales linearly with the basis dimension, we additionally perform parameter-matched experiments to decouple the effect of nonlinear diversity from that of model size~(Methods).

\begin{figure*}[htbp]
	\centering
	\includegraphics[width=0.95\textwidth]{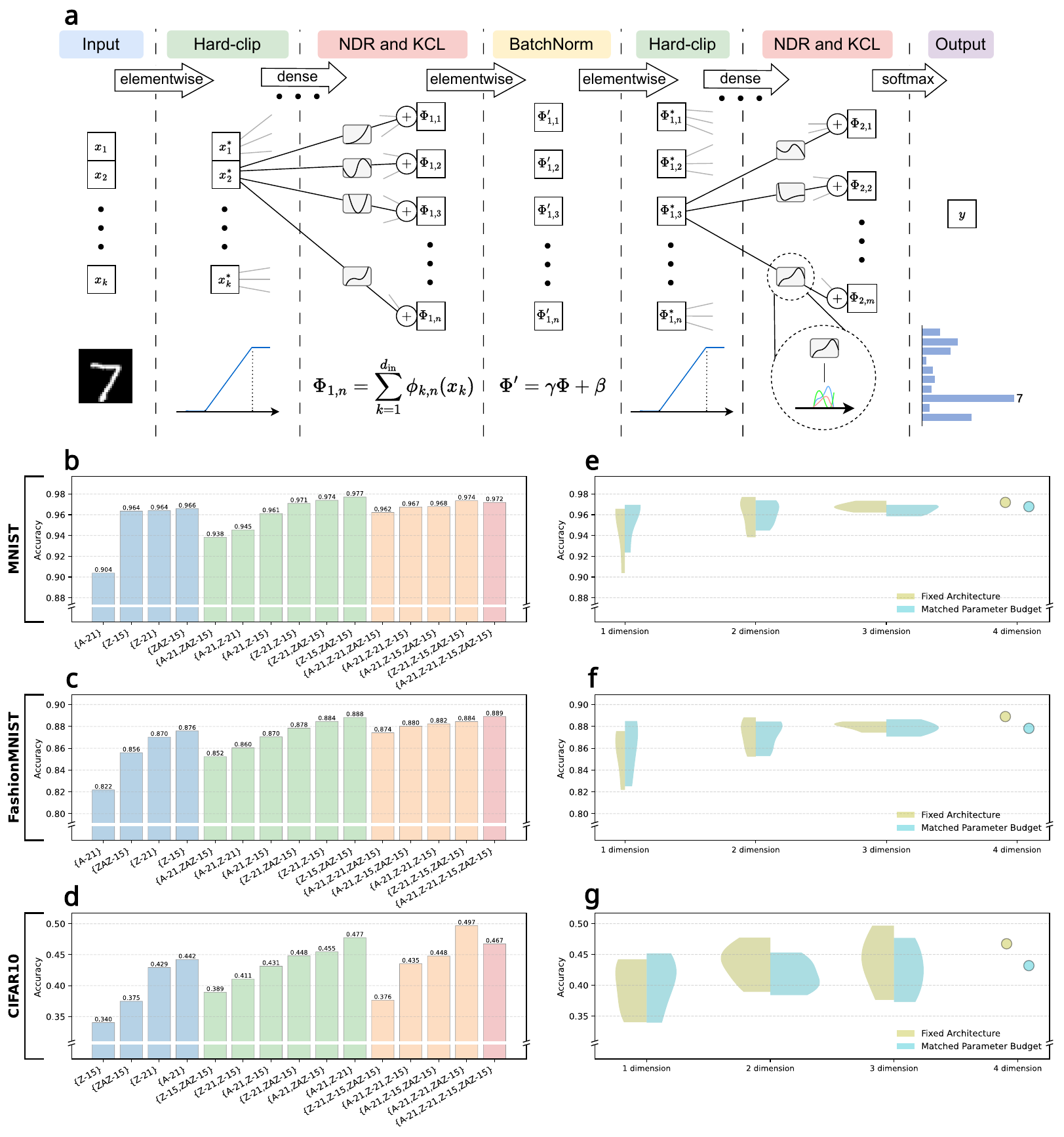}
	\captionsetup{justification=raggedright, singlelinecheck=false}
	\caption{
	\textbf{Overall pipeline for KANalogue.} Inputs are hard-clipped into the device operating range, transformed by NDR-based univariate functions and crossbar mixing, normalized during training (with BatchNorm parameters fused into weights after training), hard-clipped again if needed, and passed through subsequent analogue KAN layers before final readout. All operations—including nonlinear transformation, summation, clipping, and fused normalization—can be implemented in continuous analogue circuitry.	
	\textbf{Effect of device-derived basis-function diversity.}
		\textbf{a-c,} Classification accuracy obtained with different combinations of CMTD-based nonlinear basis functions under a fixed network architecture. Increasing the basis dimension improves peak accuracy and reduces performance variability across basis choices, particularly on more challenging datasets.
		\textbf{d-f,} Accuracy as a function of basis dimension under two experimental settings—fixed architecture and matched total parameter budget. Under parameter matching, higher basis dimensions require fewer crossbar nodes but retain comparable accuracy, indicating that performance gains arise from complementary nonlinear representations rather than increased parameter count. 
		The shorthand labels A-21, Z-21, Z15, and ZAZ-15 correspond to the CMTD device geometries shown in Fig.~\ref{fig:cmtd}.
	}
	\label{fig:pipeline_result}
\end{figure*}

Figure~\ref{fig:pipeline_result}b-d shows that increasing the basis dimension consistently improves generalization performance across all three datasets. Higher-dimensional bases not only yield higher peak accuracy, particularly on more challenging tasks, but also reduce performance variability across different combinations of basis functions. This convergence indicates that richer nonlinear representations mitigate sensitivity to the specific choice of device characteristics.

Importantly, performance gains are not solely attributable to increased parameter count. Under parameter-matched conditions (Fig.~\ref{fig:pipeline_result}e-g), models with higher basis dimensions maintain comparable accuracy despite using fewer crossbar nodes. This result demonstrates that complementary nonlinear profiles enable efficient representation.

The ablation results above demonstrate that basis-function diversity plays a central role in determining the expressive efficiency of KANalogue. Rather than relying on a single nonlinear response, combining multiple device-derived basis functions consistently improves performance within a given task. 
Based on these results, we select the best-performing basis configuration for each dataset and adopt it in all subsequent experiments. 

\subsection*{Architectural comparisons and crossbar node efficiency}
\label{sec:comparison}

To clarify the advantages of KANalogue, we compare it against two complementary reference models: 
(i) a conventional multilayer perceptron (MLP) equipped with device-based nonlinear activation functions, and (ii) the original, fully digital KAN. 
These baselines serve distinct purposes. The MLP isolates the role of nonlinear devices under a fixed network topology, while the digital KAN provides a reference for the expressive upper bound of function-based architectures unconstrained by hardware realizability.  All comparisons are conducted on MNIST, FashionMNIST, and CIFAR-10, with particular attention paid to \emph{crossbar node efficiency}, defined here as the classification accuracy achieved per analogue compute node in a crossbar implementation.

We first compare KANalogue with traditional MLPs in which digitally computed activation functions are replaced by nonlinear electronic devices, including resonant tunnelling diodes (RTDs)~\cite{ortega2021bursting} and CMTDs. In these MLP baselines, linear transformations are implemented using analogue crossbars, and each node applies a fixed device-derived nonlinearity, closely mirroring prior designs for fully analogue neural networks.  This comparison is critical: both MLPs and KANalogue employ NDR devices, but differ fundamentally in how nonlinearity is organized within the network. In an MLP, each node corresponds to a single nonlinear device with one effective degree of freedom. In KANalogue, each node represents a learned combination of multiple device-derived basis functions, resulting in a richer functional representation per node.

As shown in Table~\ref{tab:comparison_architectures}, KANalogue consistently achieves higher accuracy than device-based MLPs at the same number of nodes across all datasets. This gap widens as task complexity increases, most notably on FashionMNIST and CIFAR-10. These results demonstrate that the performance gains of KANalogue do not arise solely from the use of nonlinear devices, but from the architectural reorganization of nonlinearity enabled by the KAN formulation.

We next compare KANalogue with the original digital KAN using B-spline and Gottlieb polynomial bases~\cite{seydi2024polynomial}. Digital KANs are not constrained by device physics and therefore achieve the highest absolute accuracy~(Table~\ref{tab:comparison_architectures}). However, this performance comes at the cost of a substantially larger parameter count and reliance on digital computation.  KANalogue deliberately trades some expressive flexibility for physical realizability. Despite using a small, fixed set of device-derived nonlinearities, it achieves competitive accuracy with one to two orders of magnitude fewer parameters than digital KANs. This gap highlights a key design trade-off: digital KANs define the functional ideal, whereas KANalogue represents a hardware-grounded instantiation that preserves much of the KAN advantage under strict analogue constraints.

The advantage of KANalogue is particularly evident when accuracy is plotted as a function of crossbar node count (Fig.~\ref{fig:comparison_and_weight}a). For a fixed number of analogue crossbar nodes, KANalogue consistently outperforms both MLPs and digital KANs, indicating substantially higher crossbar node efficiency. This distinction is especially relevant for analogue hardware, where each crossbar node corresponds to physical resources such as crossbar columns, peripheral amplifiers, and routing overhead. By extracting more expressive power from each node, KANalogue reduces the required circuit scale for a target accuracy, directly translating to improved energy efficiency and throughput in fully analogue implementations.

To decouple architectural benefits from sheer model capacity, we further perform parameter-matched comparisons in which MLPs, digital KANs, and KANalogue are configured to have similar numbers of trainable parameters (Table~\ref{tab:comparison_parameters}). Under this constraint, KANalogue remains competitive across all datasets and often surpasses device-based MLPs.  Notably, even when the number of crossbar nodes is reduced to maintain a fixed parameter budget, KANalogue preserves accuracy comparable to significantly larger digital KANs. This observation reinforces that its performance advantage arises from the effective use of structured, multi-basis nonlinear representations rather than parameter scaling alone.

%
%

\begin{table*}[htbp]
	\centering
	\caption{Comparison with the same architectures on MNIST, FashionMNIST and CIFAR-10}
	\label{tab:comparison_architectures}
	\renewcommand{\arraystretch}{1.2}
	\begin{tabular}{l l c c c}
		\toprule
		\textbf{Dataset} & \textbf{Method} & \textbf{Architecture} & \textbf{\#Params} & \textbf{Accuracy (\%)} \\
		\midrule
		\multirow{7}{*}[-1.5ex]{\textbf{MNIST}} 
		& MLP~(RTD)       & [196, 64, 10]          & 13,258            & 97.17 \\ 
		& MLP~(CMTD)      & [196, 64, 10]          & 13,258            & 96.73 \\ 
		& KAN~(B-spline)   & [196, 64, 10]          & 92,288            & 97.68 \\ 
		& KAN~(Gottlieb)    & [196, 64, 10]          & 52,866            & 93.15 \\ 
		& KANalogue~(2-dim basis) & [196, 64, 10]          & 26,570            & 97.71 \\ 
		& KANalogue~(3-dim basis) & [196, 64, 10]          & 39,754            & 97.35 \\ 
		\midrule
		\multirow{7}{*}[-1.5ex]{\textbf{FashionMNIST}}
		& MLP~(RTD)       & [784, 256, 10]         & 203,530           & 88.26 \\ 
		& MLP~(CMTD)      & [784, 256, 10]         & 203,530           & 86.68 \\ 
		& KAN~(B-spline)   & [784, 256, 10]         & 1,219,584         & 89.19 \\ 
		& KAN~(Gottlieb)    & [784, 256, 10]         & 813,570           & 87.62 \\ 
		& KANalogue~(2-dim basis) & [784, 256, 10]         & 407,306           & 88.82 \\ 
		& KANalogue~(3-dim basis) & [784, 256, 10]         & 610,570           & 88.44 \\ 
		\midrule
		\multirow{7}{*}[-1.5ex]{\textbf{CIFAR-10}}
		& MLP~(RTD)       & [3072, 1024, 256, 10]  & 3,411,722         & 47.35 \\ 
		& MLP~(CMTD)      & [3072, 1024, 256, 10]  & 3,411,722         & 38.99 \\ 
		& KAN~(B-spline)   & [3072, 1024, 256, 10]  & 20,462,592        & 58.99 \\ 
		& KAN~(Gottlieb)    & [3072, 1024, 256, 10]  & 13,644,291        & 49.30 \\ 
		& KANalogue~(2-dim basis) & [3072, 1024, 256, 10]  & 6,824,714         & 47.74 \\ 
		& KANalogue~(3-dim basis) & [3072, 1024, 256, 10]  & 10,235,146        & 49.69 \\ 
		\bottomrule
	\end{tabular}
\end{table*}

\subsection*{Robustness to analogue perturbations}
\label{sec:noise}

In analogue neural systems, computational accuracy is inevitably affected by variations in physical device parameters and modeling imperfections. In KANalogue, such variability arises from two main sources: (i) device-level mismatch and fabrication-induced fluctuations, and (ii) approximation errors introduced when fitting the NDR I--V characteristics with smooth surrogate functions. Evaluating the robustness of KANalogue to these perturbations is therefore essential for assessing its practical viability.

To this end, we investigate the sensitivity of KANalogue to perturbations in the learned coefficient tensor. Specifically, we introduce multiplicative perturbations during inference of the form
\begin{equation}
	\tilde{C}_{k,j,i} = C_{k,j,i} \cdot (1 + \epsilon), 
	\quad \epsilon \sim \mathcal{N}(0, \sigma^2),
\end{equation}
where $\sigma$ denotes the magnitude of the \emph{relative coefficient perturbation}. This perturbation model captures proportional deviations in analogue coefficients, consistent with realistic device mismatch and calibration errors. Importantly, all models are trained without noise; perturbations are applied only during validation to isolate inference-time robustness.

For each relative perturbation level $\sigma$, we perform $10^4$ independent inference trials. Classification accuracy is recorded for each trial, and the worst-case accuracy across the $10^4$ realizations is reported as the main performance, reflecting conservative robustness under worst perturbation instances. In addition, the maximum and minimum accuracies observed across trials are visualized as shaded bands to illustrate the variability induced by coefficient noise. This evaluation protocol allows us to systematically characterize inference-time robustness under statistically realistic but adverse analogue perturbations.

%

\begin{figure*}[htbp]
	\centering
	\includegraphics[width=1\textwidth]{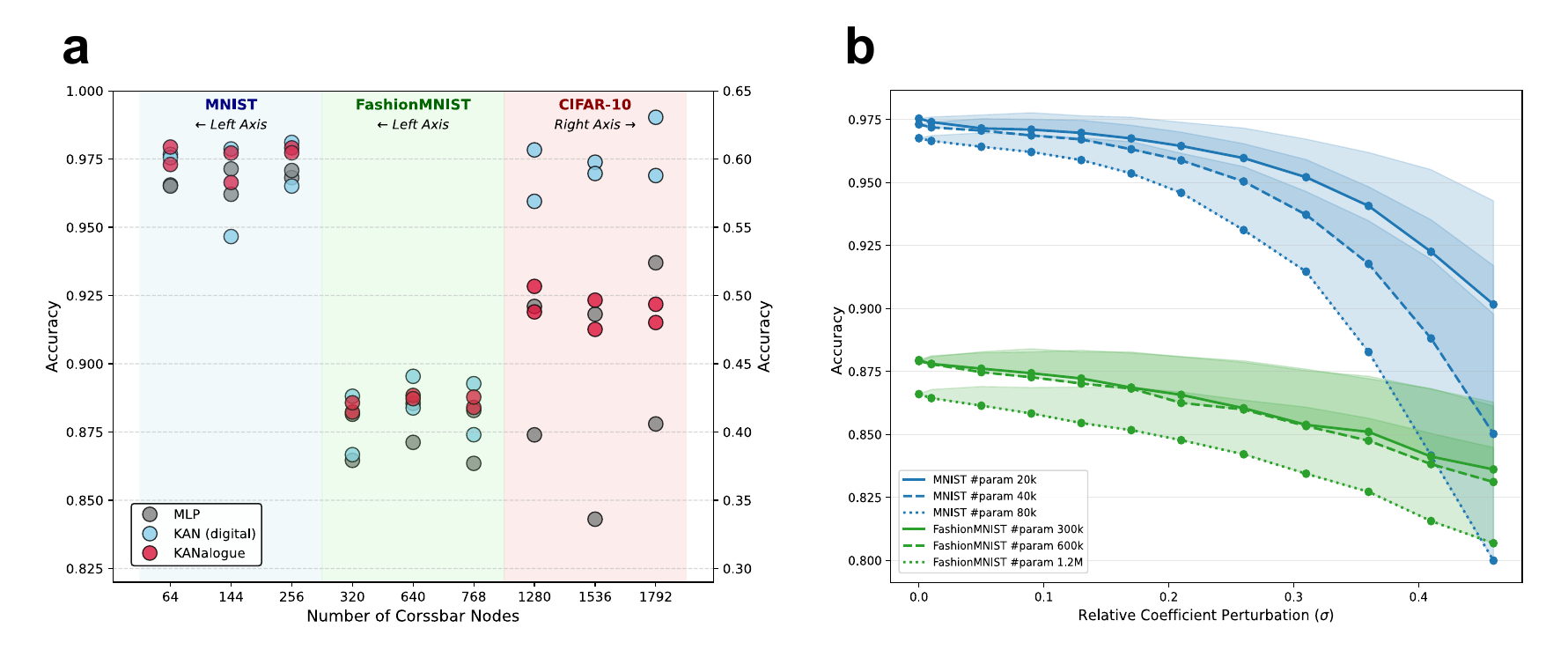}
	\captionsetup{justification=raggedright, singlelinecheck=false}
	\caption{
		\textbf{Crossbar node efficiency and robustness under coefficient disturbance.}
		\textbf{a,} Comparison of classification accuracy for MLP, KAN, and KANalogue across datasets under increasing model capacity, measured by the total number of nodes.  
		\textbf{b,} Each curve reports the worst-case classification accuracy over $10^4$ independent coefficient perturbation realizations, while the shaded bands indicate the corresponding maximum and minimum accuracies.
	}

	\label{fig:comparison_and_weight}
\end{figure*}

Figure~\ref{fig:comparison_and_weight}b reports classification accuracy as a function of the relative coefficient perturbation level for networks with varying parameter counts. As expected, increasing $\sigma$ leads to a gradual degradation in performance. However, models with more parameters exhibit markedly improved robustness: the accuracy drop becomes progressively less pronounced as the number of parameters increases. This trend indicates that increased model capacity not only enhances expressivity but also provides intrinsic error tolerance, allowing perturbations to be compensated through redundancy. 
These results suggest that scaling up KANalogue architectures offers a practical route toward robust analogue neural computation. 

\section*{Discussion}
\label{sec:discussion}

This work demonstrates that embedding device-level nonlinearities into the
KAN formalism provides a viable route toward
\emph{fully analogue} neural computation. By mapping univariate basis functions
directly onto the intrinsic current--voltage characteristics of NDR devices, KANalogue eliminates the need for digitally
implemented activations and establishes a hardware-compatible realization of
function-based learning architectures. More broadly, the results illustrate how
algorithmic structure can be co-designed with device physics to overcome a
long-standing bottleneck in analogue in-memory computing.

While CMTDs serve as a concrete instantiation in this
study, the KANalogue framework is \emph{not} limited to this specific device
technology. The only essential requirement is the availability of stable,
continuous, and sufficiently diverse univariate nonlinear responses.
Accordingly, KANalogue naturally generalizes to a broad class of NDR devices,
including Esaki diodes, resonant tunneling diodes, interband tunneling devices,
and emerging semiconductor-free or low-dimensional NDR structures. In this
sense, CMTDs should be viewed as one representative point in a much larger
design space, chosen here for their material tunability and well-behaved
nonlinear characteristics rather than as a fundamental constraint of the
architecture.

Several limitations remain. The current evaluation relies on simulated and
fitted I--V characteristics rather than large-scale fabricated arrays, and the
performance gap observed on more complex datasets reflects the finite functional
diversity available from a small set of fixed nonlinear devices. Addressing these
constraints will require both device-level advances—such as programmable or
bias-tunable NDR elements—and circuit-level co-design to manage drift, mismatch,
and operating-range constraints.

Looking forward, the KANalogue paradigm points toward a broader class of analogue
learning architectures in which physical device responses are treated as
first-class computational primitives. By grounding function approximation
directly in device physics, this approach opens a pathway toward compact,
energy-efficient, and scalable analogue neural systems that move beyond hybrid
digital--analogue pipelines and more closely align computation with the laws of the underlying hardware.

\section*{Methods}
\label{sec:methods}

%
%

\subsection*{Device Modeling and Quantum Transport Simulations}

For device modelling, we construct two-dimensional cold-metal heterostructures in which \ce{NbSi2N4} monolayers serve as metallic electrodes and a \ce{HfSi2N4} monolayer functions as the tunnelling barrier. Both armchair and zigzag crystallographic orientations are considered, with barrier thicknesses of 15~\AA{} and 21~\AA{} used to investigate geometry-dependent transport behaviour~(Supplementary Information). A vacuum layer of 20~\AA{} is applied in the out-of-plane direction to eliminate spurious interactions between periodic replicas~\cite{Bodewei2024}.

First-principles density functional theory (DFT) calculations are performed using the linear combination of atomic orbitals (LCAO) basis as implemented in QuantumATK~\cite{smidstrup2019quantumatk}. A double-zeta polarized basis set and norm-conserving FHI pseudopotentials are employed, and exchange-correlation effects are treated using the Perdew-Burke-Ernzerhof (PBE) functional within the generalized gradient approximation (GGA). To accurately reproduce the electronic band gap of the \ce{HfSi2N4} tunnelling barrier, a PBE+$U$ correction with $U=8$~eV is applied to the Hf $d$ orbitals. Self-consistent electronic structure calculations are carried out using a $24\times24\times1$ Monkhorst-Pack $k$-point mesh, a density mesh cutoff of 45~Hartree, and an energy convergence threshold of $10^{-4}$~Hartree.

Quantum transport properties are computed using density functional theory combined with the nonequilibrium Green's function (DFT-NEGF) formalism. The DFT-NEGF calculations employ a $24\times1\times184$ $k$-point grid for armchair-oriented devices and a $14\times1\times318$ grid for zigzag-oriented devices. Current-voltage characteristics are evaluated at 300~K using the Landauer formula,
\begin{equation}
I(V) = \frac{2e}{h} \int T(E,V)\,[f_L(E,V)-f_R(E,V)]\,dE,
\end{equation}
where $T(E,V)$ denotes the transmission coefficient and $f_L$ and $f_R$ are the Fermi–Dirac distributions of the left and right electrodes, respectively.

\subsection*{Evaluation Protocol}
\label{sec:methods_basis_ablation}


All basis functions are derived from spline-fitted current--voltage (I--V) characteristics of CMTDs shown in Fig.~\ref{fig:cmtd}. Four distinct nonlinear responses are selected, corresponding to different device geometries and crystallographic orientations. Each basis function is represented as a continuous univariate mapping obtained by smoothing-spline fitting of the simulated I--V data. During network training and evaluation, the shapes of these basis functions remain fixed, and only the linear mixing coefficients are optimized.

To ensure consistent operating conditions across different basis functions, the output current of each basis function is rescaled to a unified order of magnitude ($10^1$), and the input voltage is clamped to the range 
$[0, 1.4]$. This normalization and clamping procedure is applied uniformly across all experiments.

For a given experiment, a subset of the four available basis functions is selected and used jointly within each KAN layer. The number of basis functions in a configuration is referred to as the \emph{basis dimension}. All non-empty combinations of the four basis functions are evaluated, resulting in multiple configurations for each basis dimension. For each configuration, the same set of basis functions is used across all nodes and layers of the network.

Experiments are conducted on MNIST, FashionMNIST, and CIFAR-10. To account for differences in task complexity, dataset-specific network architectures are adopted. For MNIST, a single hidden-layer KANalogue network with 64 nodes is used; for FashionMNIST, a single hidden-layer network with 256 nodes; and for CIFAR-10, a two-hidden-layer architecture with 1024 and 256 nodes. Within each dataset, the network architecture is held fixed across all basis-function combinations.
Under this protocol, the total number of parameters is approximately $2.6\times10^4$ for MNIST, $4.0\times10^5$ for FashionMNIST, and $6.0\times10^6$ for CIFAR-10.

\section*{Data availability}
\label{sec:DataAvailable}
The data that support the findings of this study are publicly available datasets, including MNIST, FashionMNIST, and CIFAR-10.
The MNIST dataset is available at \url{http://yann.lecun.com/exdb/mnist/}. 
The FashionMNIST dataset is available at \url{https://github.com/zalandoresearch/fashion-mnist}. 
And the CIFAR-10 dataset is available at \url{https://www.cs.toronto.edu/~kriz/cifar.html}.

\section*{Code availability}
\label{sec:CodeAvailable}
Source code is available at \url{https://github.com/tangjr23/KANanlogue}.

\bibliography{reference/bib}

\section*{Acknowledgement}
\label{sec:acknowledgement}
The paper was supported by the National Natural Science Foundation of China under Grant Nos. T2350005 and T2541041.
The authors acknowledge the official technical support of QuantumATK provided by FermiTech (China), particularly Engineer Dong Dong, for helpful guidance and technical assistance related to the computational design.

\clearpage

\begin{appendices}

\newpage
\section{Supplementary Information}

\subsection*{Atomic structures of cold-metal tunnel diodes}

\begin{figure*}[htbp]
	\centering
	\includegraphics[width=1\textwidth]{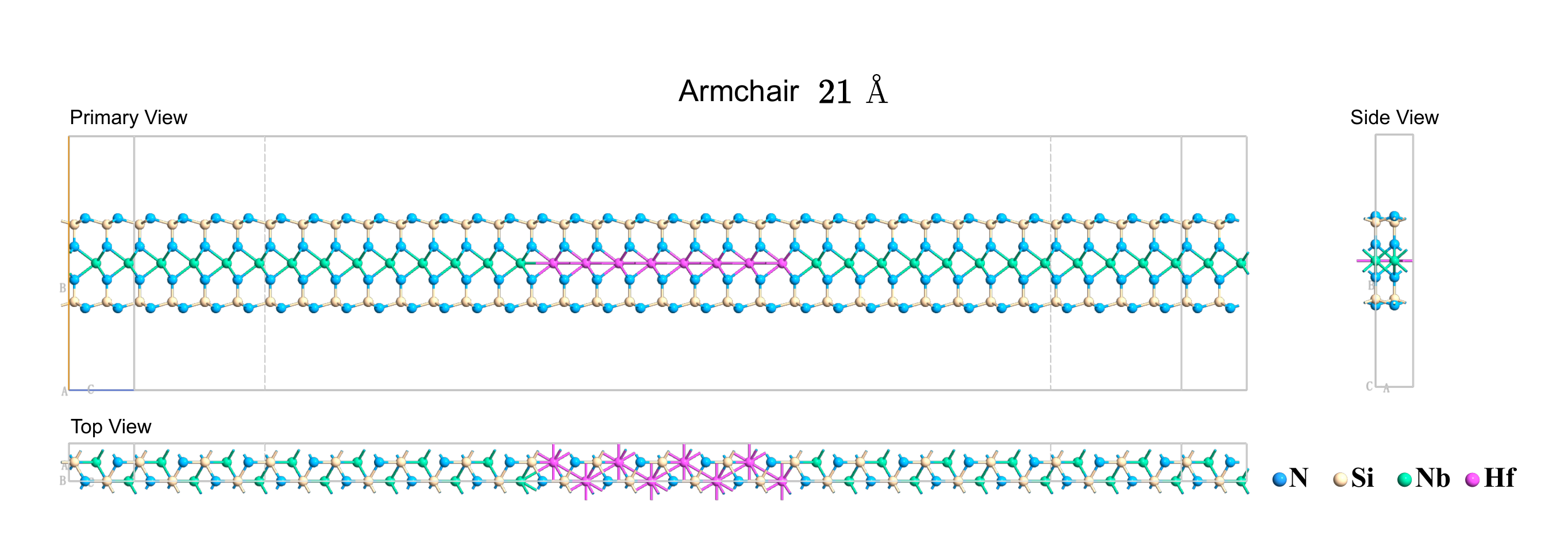}
	\captionsetup{justification=raggedright, singlelinecheck=false}
	\caption{Three-view and overview drawings of a cold-metal tunnel diode in the armchair orientation. The tunnelling barrier width is \SI{21}{\angstrom}.}
	\label{fig:a21_AtomStructure}
\end{figure*}

\begin{figure*}[htbp]
	\centering
	\includegraphics[width=1\textwidth]{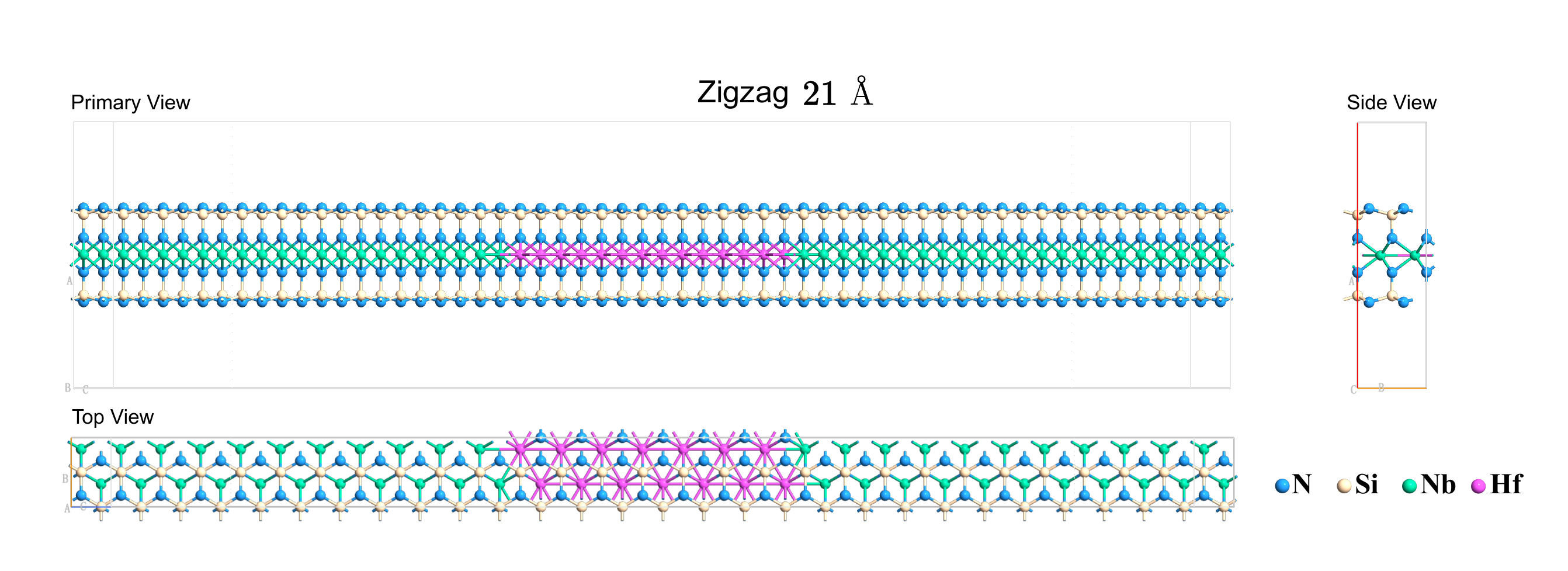}
	\captionsetup{justification=raggedright, singlelinecheck=false}
	\caption{Three-view and overview drawings of a cold-metal tunnel diode in the zigzag orientation. The tunnelling barrier width is \SI{21}{\angstrom}.}
	\label{fig:z21_AtomStructure}
\end{figure*}

\begin{figure*}[htbp]
	\centering
	\includegraphics[width=1\textwidth]{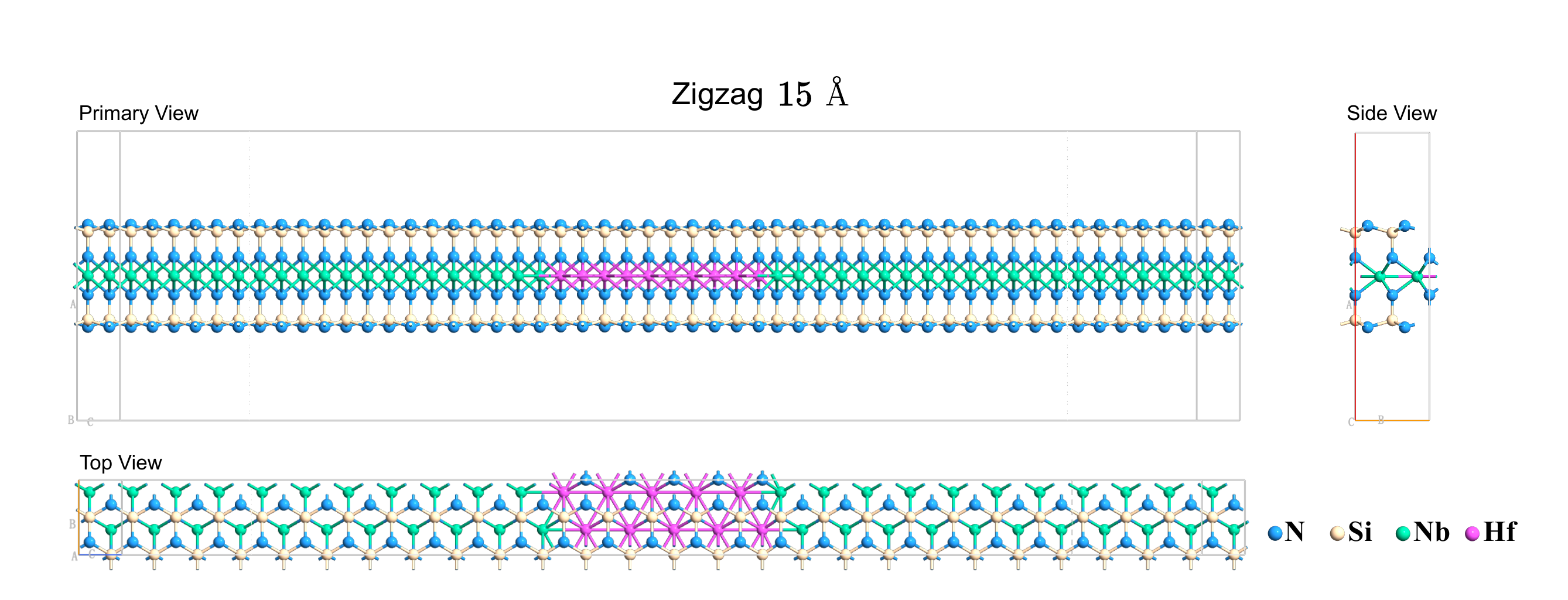}
	\captionsetup{justification=raggedright, singlelinecheck=false}
	\caption{Three-view and overview drawings of a cold-metal tunnel diode in the zigzag orientation. The tunnelling barrier width is \SI{15}{\angstrom}.}
	\label{fig:z15_AtomStructure}
\end{figure*}

\begin{figure*}[htbp]
	\centering
	\includegraphics[width=1\textwidth]{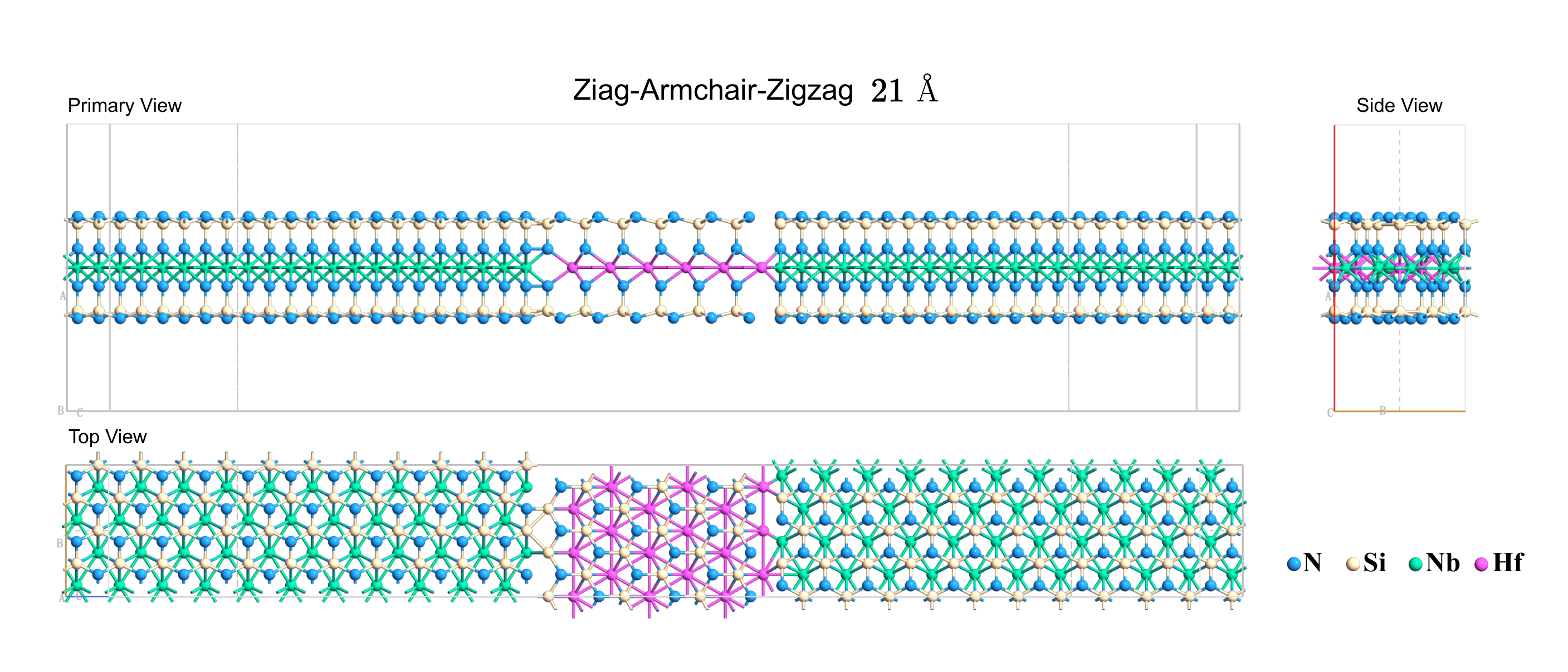}
	\captionsetup{justification=raggedright, singlelinecheck=false}
	\caption{Three-view and overview drawings of a hybrid zigzag--armchair--zigzag cold-metal tunnel diode. The tunnelling barrier width is \SI{21}{\angstrom}.}
	\label{fig:zaz21_AtomStructure}
\end{figure*}

\cref{fig:a21_AtomStructure,fig:z15_AtomStructure,fig:z21_AtomStructure,fig:zaz21_AtomStructure} present the atomic structures of the cold-metal tunnel diodes (CMTDs) investigated in this work. All devices are based on two-dimensional \ce{NbSi2N4}/\ce{HfSi2N4} lateral heterostructures, where metallic \ce{NbSi2N4} regions serve as electrodes and the \ce{HfSi2N4} segment forms the tunnelling barrier. Both armchair and zigzag crystallographic orientations are considered, together with different barrier lengths, to examine geometry- and orientation-dependent transport behaviour. 

%

\subsection*{Analysis of training and architectural choices}

We further examine the sensitivity of KANalogue to training hyperparameters and architectural choices, including batch size, learning rate, network depth, and model width. These experiments aim to assess training robustness and to clarify how scaling affects performance across datasets of increasing difficulty.

\begin{figure*}
	\centering
	\includegraphics[width=1\textwidth]{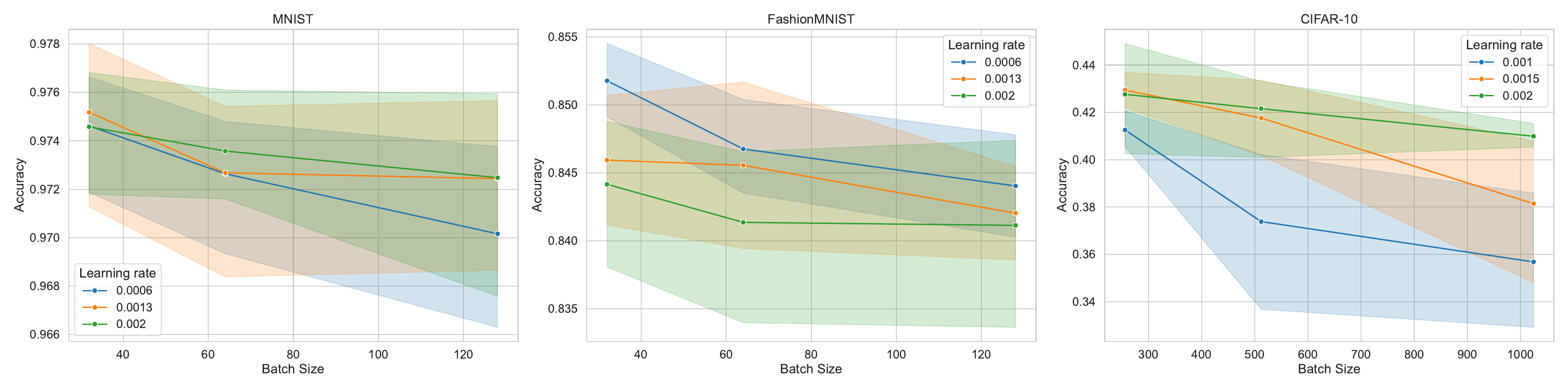}
	\captionsetup{justification=raggedright, singlelinecheck=false}
	\caption{Effect of batch size and learning rate on classification accuracy for MNIST, FashionMNIST, and CIFAR-10.}
	\label{fig:hyper_bs}
\end{figure*}

The influence of batch size and learning rate is shown in Fig.~\ref{fig:hyper_bs}. Across MNIST, FashionMNIST, and CIFAR-10, increasing the batch size leads to a mild but consistent decrease in accuracy, largely independent of the learning rate. This trend suggests that overly large batches may slightly impair generalization. Nevertheless, the overall performance variation remains limited, indicating that KANalogue is relatively robust to these hyperparameter choices.

\begin{figure*}
	\centering
	\includegraphics[width=1\textwidth]{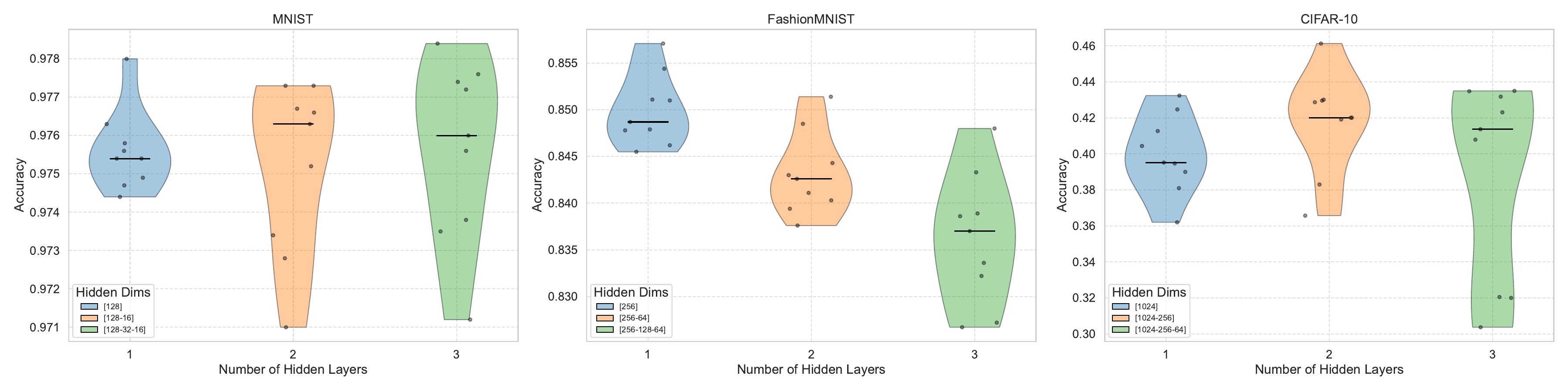}
	\captionsetup{justification=raggedright, singlelinecheck=false}
	\caption{Accuracy distributions for different numbers of hidden layers on MNIST, FashionMNIST, and CIFAR-10.}
	\label{fig:hyper_depth}
\end{figure*}

The effect of network depth is summarized in Fig.~\ref{fig:hyper_depth}. Increasing the number of hidden layers does not consistently improve accuracy and, in some cases, leads to larger variance across runs. This effect is most pronounced on CIFAR-10, where deeper architectures exhibit greater performance dispersion, suggesting increased optimization instability. These results indicate that optimal depth is task dependent and that deeper networks are not universally beneficial in the analogue KAN setting.

\begin{figure*}
	\centering
	\includegraphics[width=1\textwidth]{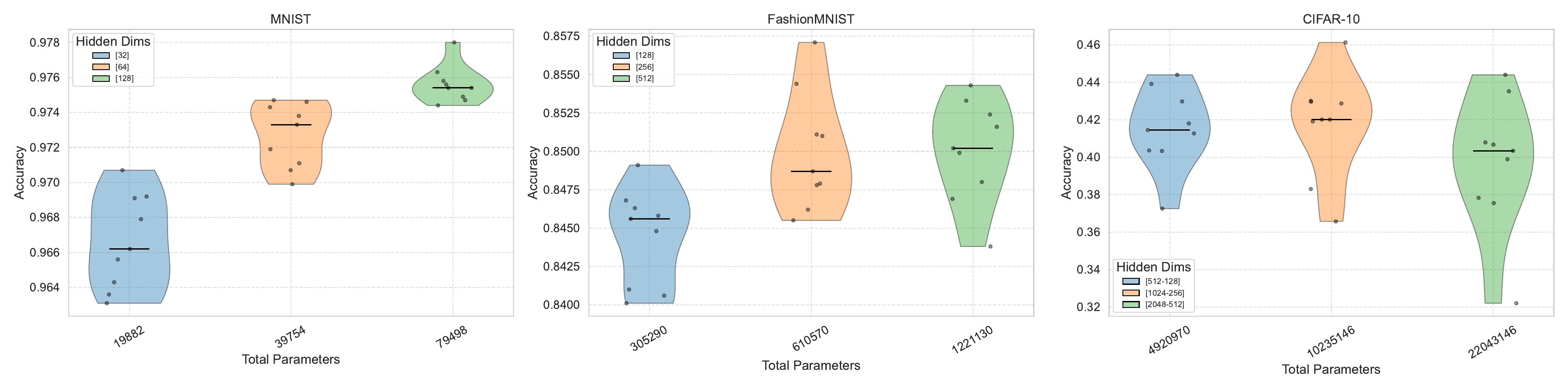}
	\captionsetup{justification=raggedright, singlelinecheck=false}
	\caption{Accuracy distributions with respect to model width under fixed depth for MNIST, FashionMNIST, and CIFAR-10.}
	\label{fig:hyper_scaling}
\end{figure*}

The impact of model width under a fixed depth is analyzed in Fig.~\ref{fig:hyper_scaling}, where results are grouped by total parameter count. Wider hidden layers generally yield higher accuracy across all three datasets; however, the gains diminish as the parameter count increases. This saturation behavior suggests that competitive performance can be achieved with moderate model sizes, without requiring aggressive width scaling.


\subsection*{Fitting, preprocessing, and normalization ablations}

\begin{figure*}
	\centering
	\includegraphics[width=1\textwidth]{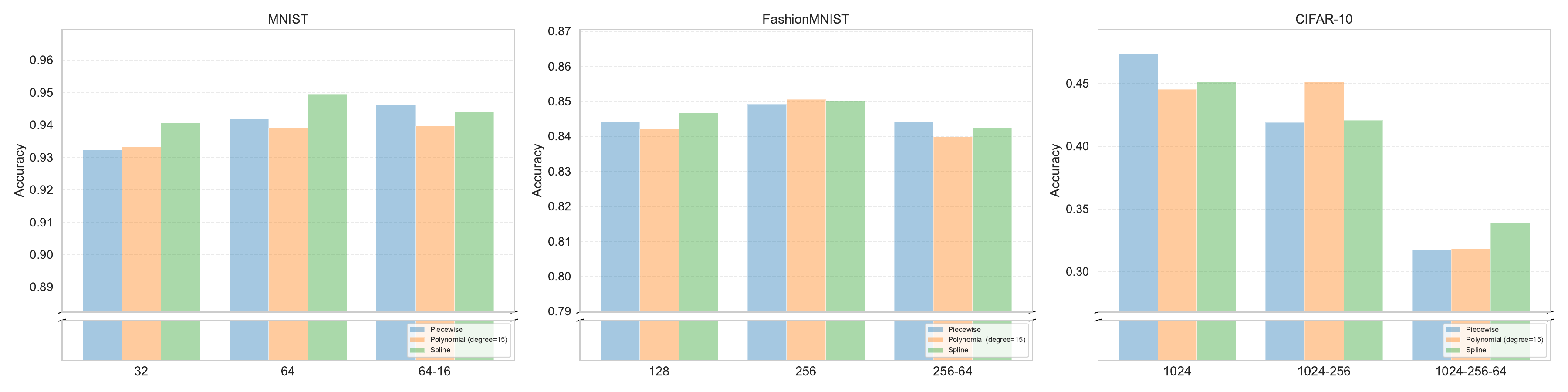}
	\captionsetup{justification=raggedright, singlelinecheck=false}
	\caption{Comparison of different fitting strategies (piecewise, polynomial, spline) on MNIST, FashionMNIST, and CIFAR-10.}
	\label{fig:hyper_fitmode}
\end{figure*}

\begin{figure*}
	\centering
	\includegraphics[width=1\textwidth]{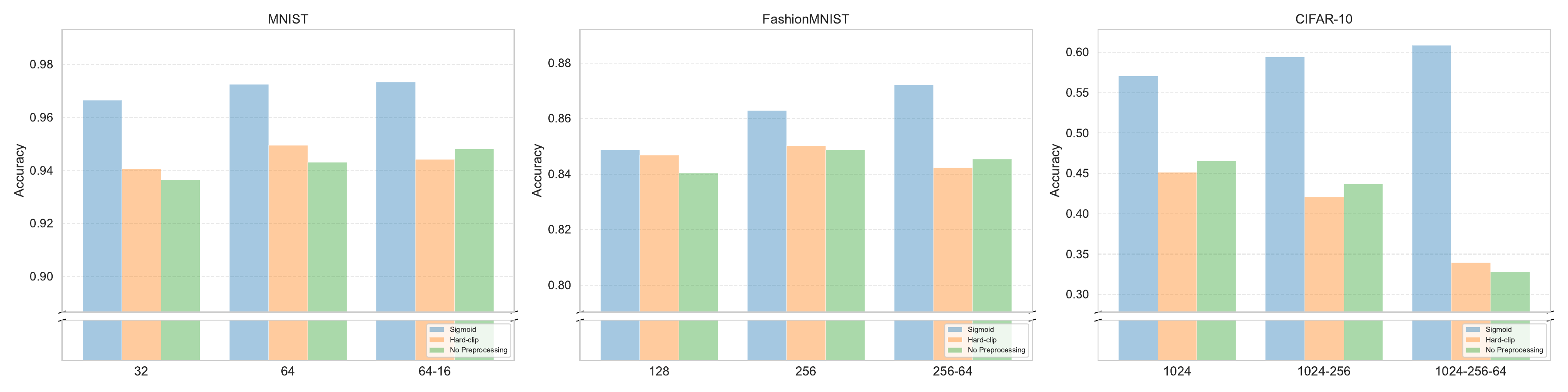}
	\captionsetup{justification=raggedright, singlelinecheck=false}
	\caption{Comparison of different input preprocessing strategies (sigmoid, hard clip, none) on MNIST, FashionMNIST, and CIFAR-10.}
	\label{fig:hyper_acti}
\end{figure*}

\begin{figure*}
	\centering
	\includegraphics[width=1\textwidth]{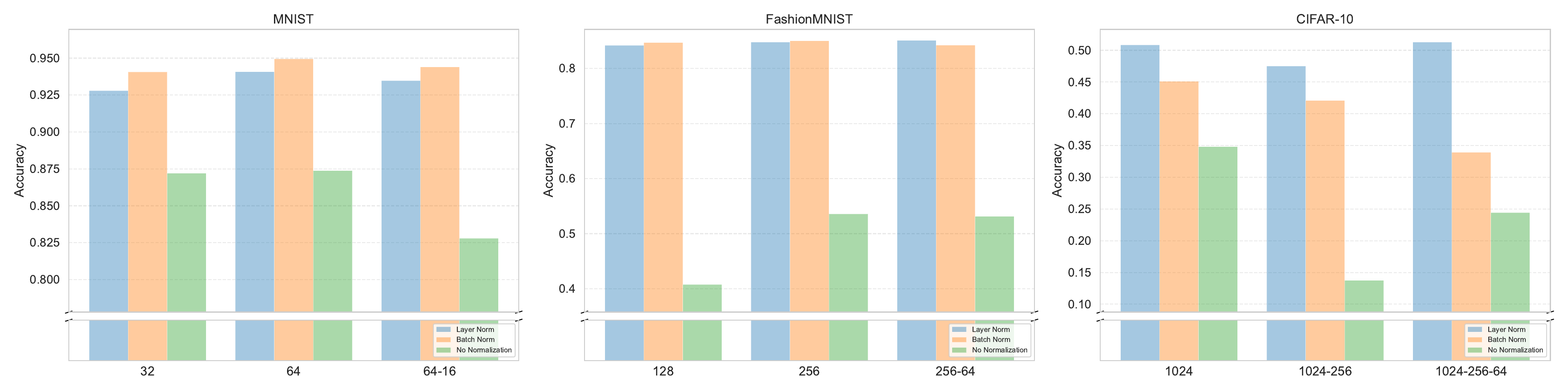}
	\captionsetup{justification=raggedright, singlelinecheck=false}
	\caption{Effect of different layer-wise normalization strategies (batch normalization, layer normalization, none) on MNIST, FashionMNIST, and CIFAR-10.}
	\label{fig:hyper_norm}
\end{figure*}

We perform additional ablation studies to examine the sensitivity of KANalogue to design choices that are relevant for analogue implementation, including the fitting strategy used to approximate device I--V characteristics, input preprocessing, and layer-wise normalization. Results are summarized in \cref{fig:hyper_fitmode,fig:hyper_acti,fig:hyper_norm}.

\cref{fig:hyper_fitmode} compares three fitting approaches for representing device I–V characteristics: piecewise fitting, high-order polynomial fitting (degree 15), and smoothing-spline fitting. Across MNIST, FashionMNIST, and CIFAR-10, all three approaches yield comparable classification accuracy under identical network configurations, indicating that KANalogue is not strongly sensitive to the specific fitting method. Among them, spline fitting consistently achieves slightly higher average accuracy and reduced variance across model sizes. This suggests that spline fitting provides a better balance between smoothness and fidelity when approximating nonlinear device responses, and is therefore adopted in the main experiments.

\cref{fig:hyper_acti} evaluates different input preprocessing strategies, including sigmoid transformation, hard clipping, and no preprocessing. Sigmoid preprocessing consistently yields the highest accuracy across all datasets, reflecting the additional nonlinearity it introduces prior to device evaluation. Hard clipping and no preprocessing achieve similar performance, particularly on simpler tasks. In practice, hard clipping remains necessary for analogue deployment to ensure that voltages remain within safe operating ranges, even when sigmoid preprocessing is not used.

\cref{fig:hyper_norm} investigates the role of normalization by comparing batch normalization, layer normalization, and no normalization. Removing normalization leads to a substantial degradation in accuracy across all datasets, highlighting its importance for stable training. For simpler tasks, batch normalization and layer normalization perform comparably, whereas for more challenging datasets such as CIFAR-10, layer normalization consistently achieves higher accuracy. Nevertheless, batch normalization remains attractive for analogue implementation because its affine parameters can be folded into static weights after training, eliminating the need for additional circuitry during inference.

%
%

\subsection*{Comparison with matched-parameter baselines}

Under parameter-matched conditions, KANalogue remains competitive across all datasets and often exceeds the performance of device-based MLPs (Table~\ref{tab:comparison_parameters}). Even with fewer crossbar nodes, it achieves accuracy comparable to much larger digital KANs, indicating that its advantage arises from structured multi-basis nonlinear representations rather than increased model capacity.

\begin{table*}[htbp]
	\centering
	\caption{Comparison with matched parameters on MNIST, FashionMNIST and CIFAR-10}
	\label{tab:comparison_parameters}
	\renewcommand{\arraystretch}{1.2}
	\begin{tabular}{l l c c c}
		\toprule
		\textbf{Dataset} & \textbf{Method} & \textbf{Architecture} & \textbf{\#Params} & \textbf{Accuracy (\%)} \\
		\midrule
		\multirow{7}{*}[-1.5ex]{\textbf{MNIST}}
		& MLP~(RTD)       & [196, 128, 10]         & 26,506            & 97.63 \\ 
		& MLP~(CMTD)      & [196, 128, 10]         & 26,506            & 97.36 \\ 
		& KAN~(B-spline)   & [196, 20, 10]          & 28,840            & 96.85 \\ 
		& KAN~(Gottlieb)    & [196, 32, 10]          & 26,434            & 96.71 \\ 
		& KANalogue~(2-dim basis) & [196, 64, 10]          & 26,570            & 97.38 \\
		& KANalogue~(3-dim basis) & [196, 42, 10]          & 26,092            & 96.96 \\
		\midrule
		\multirow{7}{*}[-1.5ex]{\textbf{FashionMNIST}}
		& MLP~(RTD)       & [784, 512, 64, 10]     & 435,402           & 88.48 \\ 
		& MLP~(CMTD)      & [784, 512, 64, 10]     & 435,402           & 84.61 \\ 
		& KAN~(B-spline)   & [784, 80, 10]          & 444,640           & 88.60 \\ 
		& KAN~(Gottlieb)    & [784, 128, 10]         & 406,786           & 85.31 \\ 
		& KANalogue~(2-dim basis) & [784, 256, 10]         & 407,306           & 88.44 \\ 
		& KANalogue~(3-dim basis) & [784, 170, 10]         & 405,460           & 88.65 \\ 
		\midrule
		\multirow{7}{*}[-1.5ex]{\textbf{CIFAR-10}}
		& MLP~(RTD)       & [3072, 2048, 512, 10]  & 7,347,722         & 48.93 \\ 
		& MLP~(CMTD)      & [3072, 2048, 512, 10]  & 7,347,722         & 40.02 \\ 
		& KAN~(B-spline)   & [3072, 300, 256, 10]   & 7,006,720         & 58.97 \\ 
		& KAN~(Gottlieb)    & [3072, 512, 256, 10]   & 6,827,523         & 53.11 \\ 
		& KANalogue~(2-dim basis) & [3072, 1024, 256, 10]  & 6,824,714         & 45.31 \\ 
		& KANalogue~(3-dim basis) & [3072, 682, 256, 10]   & 6,819,592         & 47.70 \\
		\bottomrule
	\end{tabular}
\end{table*}

\end{appendices}

\end{document}